\documentclass[pra,aps,twocolumn,showpacs,superscriptaddress,amsmath,amssymb]{revtex4-1}
\usepackage{graphicx}
\input{epsf}
\begin{document}
\epsfverbosetrue
\def\la{{\langle}}
\def\ra{{\rangle}}
\def\vep{{\varepsilon}}
\newcommand{\beq}{\begin{equation}}
\newcommand{\eeq}{\end{equation}}
\newcommand{\beqa}{\begin{eqnarray}}
\newcommand{\eeqa}{\end{eqnarray}}
\newcommand{\da}{^\dagger}
\newcommand{\wh}{\widehat}
\newcommand{\os}[1]{#1_{\hbox{\scriptsize{osc}}}}
\newcommand{\cn}[1]{#1_{\hbox{\scriptsize{con}}}}
\newcommand{\sy}[1]{#1_{\hbox{\scriptsize{sys}}}}
\newcommand{\BEC}{}
\newcommand{\q}{\quad}
\newcommand{\lm}{\lambda}
\newcommand{\lt}{\tilde{\lambda}}
\newcommand{\e}{\tilde{E}}
\newcommand{\W}{{\bf \Psi}}
\newcommand{\U}{{\bf u}}
\newcommand{\V}{\hat{V}}
\newcommand{\F}{{\bf\Phi}}
\newcommand{\spm}{\hat{S}}
\newcommand{\br}{\\ \nonumber}

\title{Self-intersecting Regge trajectories in multi-channel scattering}
\author{D. Sokolovski*}
\affiliation{Department of Chemical Physics, University of the Basque Country, Leioa, Spain.}
\affiliation{IKERBASQUE, Basque Foundation for Science, 48011, Bilbao, Spain.}
\author {Z. Felfli}
\author{A. Z. Msezane}
\affiliation{Department of Physics and Center for Theoretical Studies of Physical Systems, Clark Atlanta University, Atlanta, Georgia 30314, USA}

\begin{abstract}
{*Corresponding author: dgsokol15@gmail.com, Ph.+34 633557019}
\vspace{0.8cm}
\newline
We present a simple direct method for calculating Regge trajectories for a multichannel scattering problem. The approach is applied to the case of two coupled Thomas-Fermi type potentials, used as
a crude model for electron-atom scattering below the second excitation threshold.
It is shown that non-adiabatic interaction may cause formation of loops in Regge trajectories.
The accuracy of the method is tested by evaluating resonance contributions to elastic and inelastic integral cross sections.

\end{abstract}

\date{\today}

\maketitle

\vskip0.5cm
%
\section{Introduction}
There has long been interest in resonance effects which arise when collision partners form a long-lived intermediate complex \cite{Tayl}. Recently, this interest was reinforced by experimental progress in cold atomic and molecular collisions  (see, for example, \cite{Cost}). 
An isolated resonance can be associated with a pole of the scattering matrix either in the complex energy (CE) or the complex angular momentum (CAM) plane. Two types of poles are closely related
and contain essentially the same amount of information.
However,  since observables of interest such as integral (ICS) and differential (DCS) cross sections are given by partial wave sums (PWS) over total angular momentum $J$, the CAM (Regge) poles prove to be more convenient for identifying and quantifying resonance effects.  Transforming a PWS into a sum of integrals, {\it e.g.}, by means of the Poisson sum formula, and then evaluating contributions from Regge poles often allows to account for the resonance pattern observed in a DCS \cite{N1}-\cite{S2}  or an ICS \cite{Mac}-\cite{SA}. 

Applications of the CAM approach range from elastic collisions of atoms with protons \cite{Mac} and electrons \cite{e},\cite{AZM} to atom-diatom chemical reactions \cite{S2},\cite{S1}. Numerous techniques have been proposed for determination of Regge pole positions and residues in single channel potential scattering
(for a review see \cite{Conn}),
 among others direct solution of the Schroedinger equation \cite{Burke} (see also \cite{e}), or of the corresponding non-linear Milne equation \cite{Th1,Th2} for complex values of $J$. For  realistic reactive systems the number of open channel is large, root search in the complex $J$ plane is not viable, and one has to
resort to Pade' reconstruction of $S$-matrix elements \cite{S3}.
This leaves a class of systems with a relatively few channels, for which many of 
the single channel techniques do not work, and yet one wishes to avoid the use of Pade' approximants.

The purpose of the present paper is to propose a direct method for calculating CAM poles positions and residues for such systems.
These include, among others,  inelastic and reactive systems at low energies, Feshbach resonances, collisions involving two-level atoms \cite{Nik}, and spin flip scattering \cite{flip}.
We will also look for the evidence of non-adiabatic effects in the behaviour of Regge trajectories.
For recent efforts in this direction we refer the reader to Ref.\cite{Th3}, where the amplitude-phase method of Refs.\cite{Th1,Th2} has been extended to Dirac electrons.
The rest of the paper is organised as follows. In Sect. II we give a brief description of the method
which generalises the approach of Ref. \cite{Burke} to a multichannel case. In Sect. III we consider a two-channel case designed to mimic electron-atom scattering below
the second excitation threshold. Section IV considers the same problem in the adiabatic approximation. In Sect. V we use the obtained pole positions and residues to evaluate resonance contributions to elastic and inelastic integral cross sections. Section VI contains our conclusions.

\section{Direct calculation of Regge pole positions and residues}
Consider a time-independent scattering problem described by $N$ coupled radial equations
[we set to unity the particle's mass, $\mu=1$, and choose $\hbar=1$, thus converting to atomic units (a.u,)],
\begin{eqnarray}\label{1}
\{[-\partial_r^2/2+J(J+1)/2r^2-E]\hat{I}
+\V(r) \} \W(r)=0
\end{eqnarray}
where $E$ is the energy, $J$ is the total angular momentum,  $\hat{I}$ is the unit matrix, and $\V(r)$ is an $N\times N$ hermitian potential matrix, such that
\begin{eqnarray}\label{2}
lim_{r\to \infty}\V(r)=diag(V_1,V_2,..., V_{N-1},0).
\end{eqnarray}
We will assume that the constant values $V_n$ are arranged in such a way that $V_1\ge V_2\ge...\ge V_{N-1}$,
and require that a solution of Eq.(\ref{1}),
given by a complex vector
$\W(r)=[\Psi_1,\Psi_2,...,\Psi_n]^T$, is regular at the origin,
\begin{eqnarray}\label{3}
lim_{r\to 0}\Psi_n(r)=0, \quad n=1,2,...,N.
\end{eqnarray}
Furthermore, we are interested in scattering solutions $\W_m$, $m=1,2,...N$, which for large $r's$ contain an incoming wave 
in only the $m$-th channel.
Assuming that the potential $\V(r)$ reaches its asymptotic form (\ref{2}) sufficiently rapidly,
as $r\to \infty$ for the channel wavefunctions we have
\begin{eqnarray}\label{4}
{(\W_m)_n}
\approx (\pi k_nr/2)^{1/2}[\delta_{nm}H^{(2)}_{\lm}(k_nr)+
\\
\nonumber S_{nm} (E,\lambda) H^{(1)}_{\lm}(k_nr)], \q n,m=1,2...N 
\end{eqnarray}
where  $H^{(1)}_{\lm}(z)$
and $H^{(2)}_{\lm}(z)$ are the Hankel functions of the first and second kind, respectively, 
$\lambda \equiv J+1/2$, and the asymptotic wave vector $k_n$ is given by
\begin{eqnarray}\label{4a}
k_n\equiv \sqrt{2(E-V_n)},\q n=1,2,...N.
\end{eqnarray}
For now we will assume that all channels are open, $E>V_1$ so that all $k_n$ are real valued.
For each real value of the energy $E$ we wish to find complex value(s) $\lambda=\bar{\lambda}(E)$ ,
such that the $S$-matrix elements $S_{nm}$ would diverge, $S_{nm}(E,\bar{\lambda}(E))=\infty$.
Thus, the asymptotic of the corresponding solution of Eq.(\ref{1}) (Regge state) will contain only  outgoing waves
generated by the emissive complex centrifugal potential $[\bar{\lambda}(E)^2-1/4]/2r^2$.

In order to obtain Regge trajectory(s) $\bar{\lambda}(E)$ we integrate Eq.(\ref{1}) for an arbitrary 
complex value of the angular momentum $J$ ($\lambda$) sufficiently far into the asymptotic region, evaluate the $S$-matrix elements, and repeat the procedure until a (Regge) pole of the $S$-matrix  is found \cite{e,Burke}. With possible applications of the theory to electron-atom collisions in mind, we consider a potential matrix which has a Coulomb singularity at the origin,
so that its Taylor expansion takes the form
\begin{eqnarray}\label{5}
\V(r)=r^{-1}\sum_{j=0}^{\infty} \V_jr^j.
\end{eqnarray}
As in the one-channel case \cite{e,Burke} the singularity of $\V(r)$ prevents imposing 
the boundary condition (\ref{3}) directly at $r=0$.  Following \cite{Burke}
we represent the solution of Eq. (1)as a power series,
\begin{eqnarray}\label{6}
\W(r)=r^{J+1}\sum_{j=0}^{\infty} \U_jr^j.
\end{eqnarray}
 where $\U_j$, $j=0,1,...$, are constant vector coefficients satisfying recursion relations
 \begin{eqnarray}\label{7}
\sum_{j=1}^{\infty} \U_{j+1}j(j+1) r^{j}
+2(J+1)\sum_{j=0}^{\infty} \U_{j+1}(j+1) r^{j}\\
\nonumber
+\sum_{j=0}^{\infty}(\sum_{l=0}^{j}\hat{K}_l \U_{j-l})r^l =0
\end{eqnarray}
with $\hat{K}_l \equiv \V_l+E\hat{I}\delta_{l,1}$,  whose explicit solution reads 
\begin{eqnarray}\label{8}
\U_{j+1}=-[2(J+1)(j+1)+j(j+1)]^{-1}\sum_{l=0}^{j}\hat{K}_l \U_{j-l}, \q j\ge0.\q \q
\end{eqnarray}
The recursion scheme is initialised by specifying the so far undefined initial vector $\U_0$.
This can be chosen in $N$ different ways, e.g., (the last subscript indicates the component 
of the vector $\U_{0}^{m} $)
 \begin{eqnarray}\label{9}
(\U_{0}^m)_n=\delta_{nm} \q n,m=1,2,...,N
\end{eqnarray}
to yield $N$ linearly independent 
 solutions,
whose values at some $r_0$ for a suitably chosen $j_{max}$.
 \begin{eqnarray}\label{10}
\F_m(r_0)=\sum_{j=0}^{j_{max}} \U_j^m r_0^{J+j+1},\q\q\q\q\q\q\q\q\q\q\\
\nonumber 
\F'_m(r_0)=\sum_{j=0}^{j_{max}} (J+j+1)\U_j^m r_0^{J+j}, \q m=1,2,..N,
\end{eqnarray}
provide $N$ sets of initial conditions for Eq.(\ref{1}).
This can now be integrated numerically (a NAG integrator \cite{NAGINT} is used in this work) 
to a sufficiently large $r$ where
\begin{eqnarray}\label{11}
(\F_m)_n
\approx (\pi k_nr/2)^{1/2}[\spm^-_{mn}H^{(2)}_{\lm}(k_nr)+
\br
 \spm^+_{mn}  H^{(1)}_{\lm}(k_nr)], \q n,m=1,2...N 
\end{eqnarray}
and  $\spm^-(E,\lambda)$ and $\spm^+(E,\lambda)$ are constant matrices, to be determined numerically.
 The physical scattering states (\ref{4}) are linear combinations of 
$\F_m$, 
\begin{eqnarray}\label{12}
\W_k=\sum_m A_{km}\F_m, \q k=1,2,...,N,
\end{eqnarray}
with  $A_{km}$ chosen so that 
the coefficients multiplying the Hankel functions of the 
second kind (incoming waves) add up to $\delta_{mn}$ 
i.e., $\hat{A}=(\spm^-)^{-1}$.
As a result, for the $S$-matrix we have
\begin{eqnarray}\label{14}
S_{nn'}= [(\spm^-)^{-1}\spm^+]_{nn'}.
\end{eqnarray}
It is readily seen that the $S$-matrix elements diverge if and only if $\spm^-$ is singular, so that the condition 
for a Regge pole at a (real) energy $E$ reads
\begin{eqnarray}\label{15}
\Delta(E,\lambda)\equiv \det\spm^-=0.	
\end{eqnarray}
Starting with a reasonable initial guess for $\lambda$ and recalculating the l.h.s. of Eq.(\ref{15})  in each step, one can use a standard routine
for finding zeroes of $\Delta(E,\lambda)$  (a NAG root finder \cite{NAGROOT} is used in this work) to determine the accurate pole position  $\bar{\lambda}(E)$. The residues 
 \begin{eqnarray}\label{16}
\rho_{nn'}(E)\equiv \lim_{\epsilon \to 0} \epsilon S_{nn'}(E,\bar{\lambda}+\epsilon).	
\end{eqnarray}
are readily obtained by integrating Eq.(\ref{1}) for a value of $\lambda$ close to $\bar{\lambda}(E)$
and taking the limit (\ref{16}).
Finally, in the case some of the channels are closed, with corresponding $k_n$'s in Eq.(\ref{4a}) purely imaginary, equation (\ref{15}) applies, and ensures that the Regge state does not have components which grow exponentially as $r\to \infty$. 
 After this brief summary, in the next Sections we apply the method  to a model two-channel 
$(N=2)$ problem.
\section{The model: two coupled Thomas-Fermi type potentials}

Next we consider a two-channel scattering problem with a potential matrix defined
by 
\begin{eqnarray}\label{17}
V_{22}(r)=-\frac{Z}{r(r+a)(r^2+b)}
\end{eqnarray}
\begin{eqnarray}\label{17a}
V_{11}(r)=V_{22}(r)+\Delta V,
\end{eqnarray}
and 
\begin{eqnarray}\label{18}
V_{12}(r)=V_{21}(r)=\alpha\exp[-(r-r_i)^2/\Delta r^2].
\end{eqnarray}
The potential shown in Fig.1 can be seen as a crude model for an inelastic electron-atom collision below
the threshold of the second inelastic channel.
The diagonal terms $V_{11}$ and $V_{22}$ are two similar Thomas-Fermi type potentials, representing the interaction between an electron and an atom in the first excited and the ground state, respectively. The constant $\Delta V$ is the excitation energy, and the interaction between the channels  occurs in the outer layer of the atom, $r_i-\Delta r \lesssim r \lesssim r_i+\Delta r$.
 \begin{figure}[ht]
\centerline{\includegraphics[width=8cm, angle=-0]{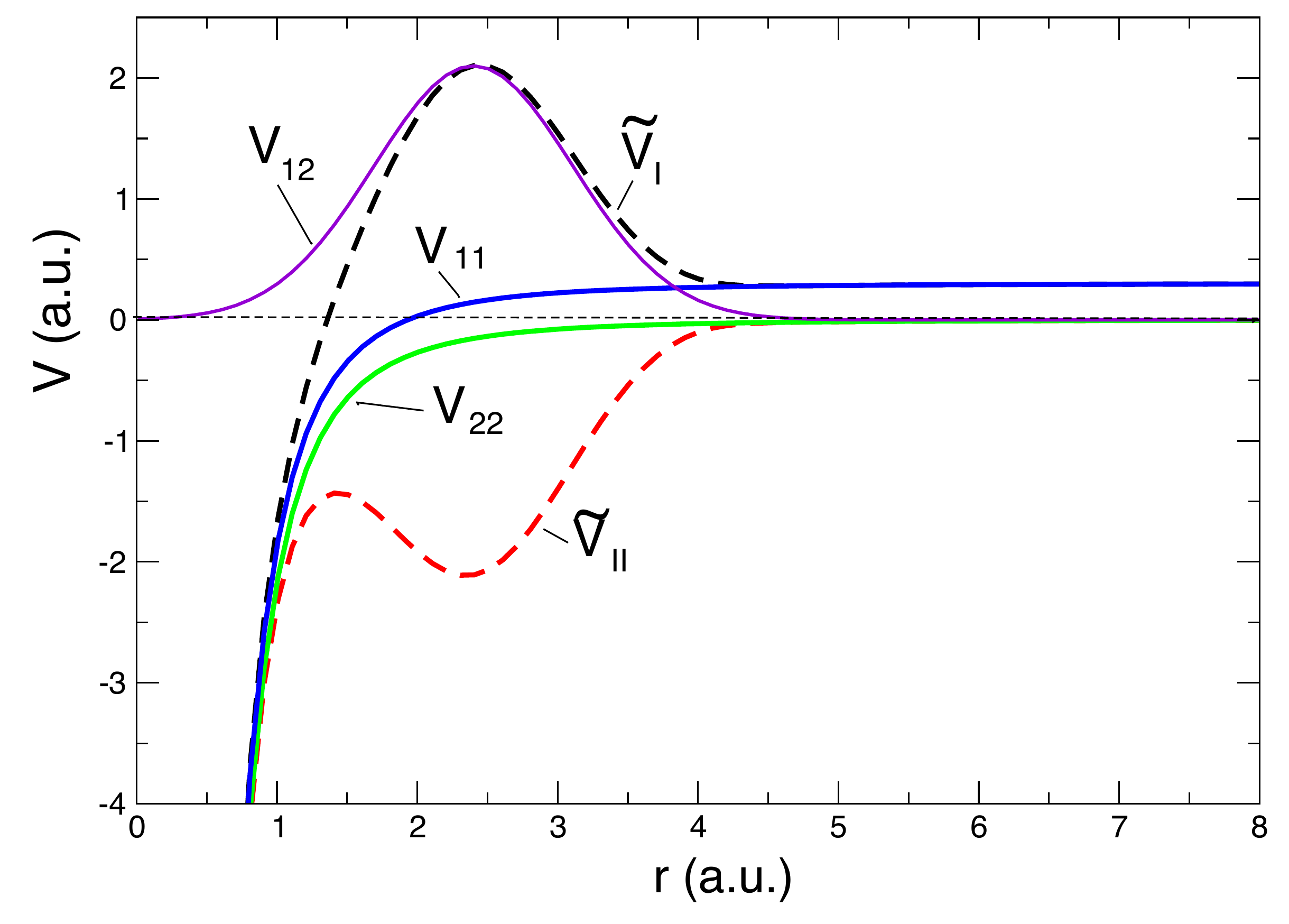}}
\caption{ (Color online) Elements of the potential matrix vs. $r$ (solid).
Also shown by dashed lines are the adiabatic potential curves (\ref{19}).}
\label{fig:1}
\end{figure}
Regge trajectories (curves $Im \bar{\lambda}$ vs. $Re \bar{\lambda}$) are shown in Fig. 2 for
arbitrary $Z=54$, $a=0.0125$ a.u., $b=1.5874$ a.u. and $\Delta V=0.3$ a.u. For uncoupled channels ($V_{12}=V_{21}=0$),
there are two Regge trajectories, $\bar{\lambda}_{I}(E)$ and  $\bar{\lambda}_{II}(E)$.
Since the two potentials only differ by a constant shift $\Delta V$, $\bar{\lambda}_{I}(E)=\bar{\lambda}_{II}(E+\Delta V)$, and the two Regge trajectories in Fig. 2 coincide, as shown by the dot-dashed line. Interaction between the channels removes the degeneracy and yields two distinct
trajectories as shown in Fig.2 for
 $\alpha=1.5$ a.u.,   $\Delta r=1$ a.u., and  
 $r_i=2.4$ a.u. (solid).
  \begin{figure}[ht]
\centerline{\includegraphics[width=8cm, angle=-0]{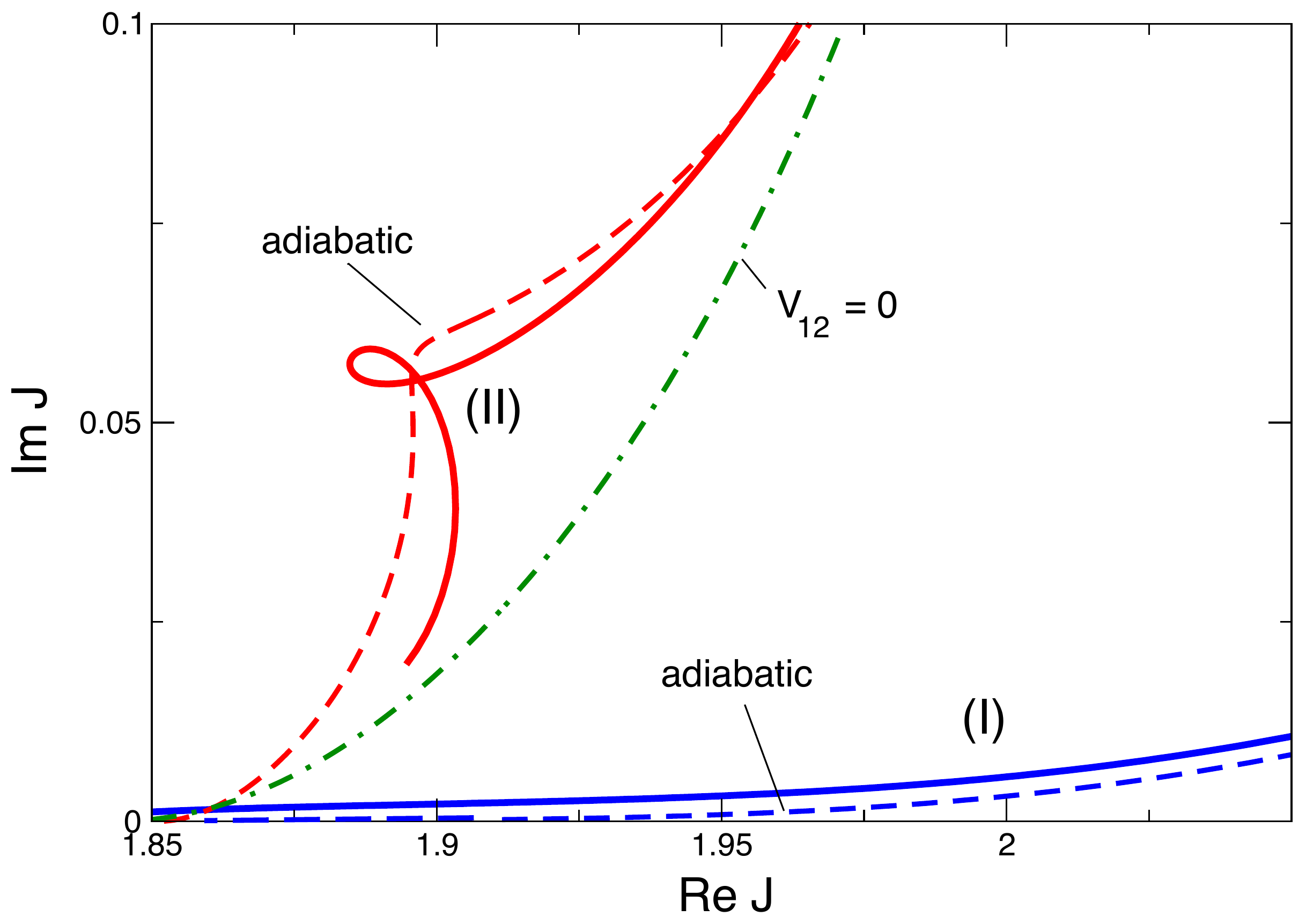}}
\caption{ (Color online) Regge trajectories
$Im J$ vs. $Re J$
 for the potential in Fig. 1 (solid) shown for $E>\Delta V$.
Also shown are Regge trajectories in the adiabatic approximation (\ref{19}) (dashed)
and the two coinciding Regge trajectories for $V_{12}=V_{21}=0$ (dot-dashed)}
\label{fig:1}
\end{figure}
The trajectory labelled $(I)$ exhibits a type of behaviour often seen in potential scattering \cite{Mac,e,SA}. As the energy increases, so does $\bar{\lambda}_{I}(E)$, and the trajectory 
curves away from the real $\lambda$-axis in a smooth manner.
The second trajectory, labelled $(II)$  leaves the real axis much more rapidly, and shortly 
thereafter intersects itself, describing a loop in the first quadrant of the CAM plane.
This behaviour, to our knowledge not observed in one-channel scattering problems, 
is one of our central results. In the next Section we demonstrate it to be a consequence of non-adiabatic effects.
 \section{Adiabatic correspondence}
Further insight can be gained by considering the two Regge trajectories in the adiabatic approximation.
Diagonalising the potential matrix $\hat{V}(r)$ for each value of $r$ yields two adiabatic curves,
 \begin{eqnarray}\label{19}
\tilde{V}_{I,II}(r)=\frac{V_1+V_2}{2}\pm\frac{\sqrt{(V_1-V_2)^2+4V_{12}}}{2},
\end{eqnarray}
shown in Fig.1 by dashed lines.
The first $(I)$ curve acquires an additional barrier, while the second one $(II)$ has an additional 
well, both roughly proportional to $\pm V_{12}(r)$.
 Neglecting non-adiabatic coupling, i.e., replacing in Eq.(\ref{1}) 
 \begin{eqnarray}\label{20}
 \hat{V}(r) \to diag[\tilde{V}_{I}(r),\tilde{V}_{II}(r)], 
\end{eqnarray} 
 yields two uncoupled equations and two adiabatic Regge trajectories shown in Fig.2 by dashed lines. The first adiabatic trajectory is close to the exact trajectory $(I)$ in Fig.2,
 whose imaginary part grows slowly with the energy, since the corresponding metastable state is stabilised by the effective barrier shown in Fig.1. The second exact trajectory $(II)$ corresponds to the adiabatic trajectory for the barrierless potential $\tilde{V}_{II}(r)$. Comparing these two Regge trajectories suggests that self-intersection of the exact curve in Fig.2 is caused by the non-adiabatic transitions not taken into account by the approximation (\ref{19})-(\ref{20}).
   \begin{figure}[ht]
\centerline{\includegraphics[width=7cm, angle=-0]{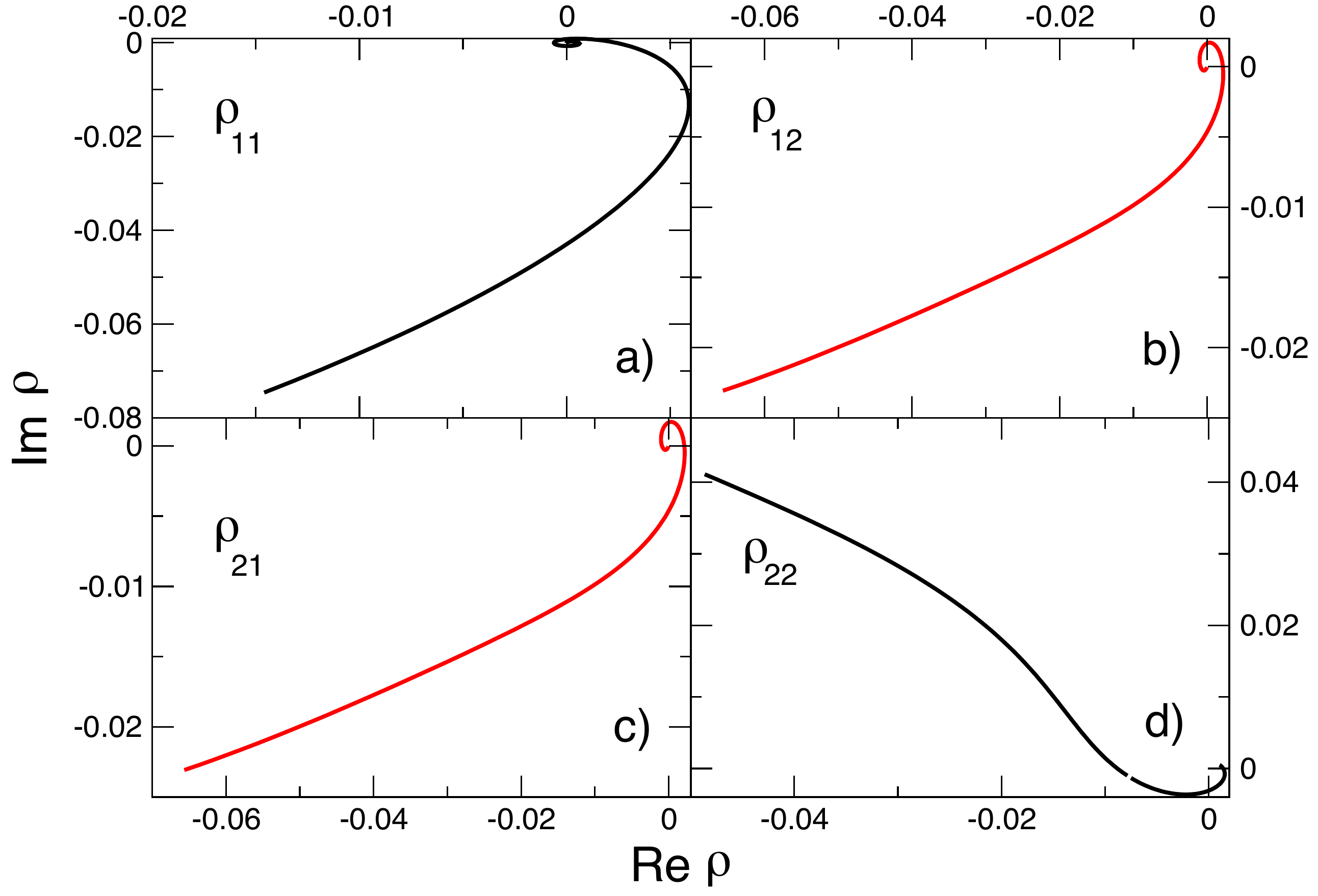}}
\caption{ (Color online) Residue trajectories, $Im \rho_{ij}$ vs. $Re \rho_{ij}$
for (a) i=1, j=1; (b) i=1, j=2; (c) i=2, j=1; (d) i=2, j=2, for the trajectory $(I)$ in Fig.2}
\label{fig:1}
\end{figure}

\section{Integral cross sections.}
One important application of the CAM theory is in the identification and quantitative analysis of resonance patterns which occur in
elastic, inelastic and reactive integral cross sections (ICS) \cite{Mac}-\cite{SA}. A resonance is likely to affect an ICS at an energy $E>0$ for which the corresponding  Regge trajectory approaches a real integer value of the angular momentum \cite{Mac},\cite{e}. 
(The requirement is readily understood if one recalls that at negative energies the condition for a true bound state is 
that $J$ take a 'physical' integer value.) This condition is satisfied for the trajectory (I) in Fig. 2 which approaches 
$J=2$ for $E\approx 1.46$ a.u. Next we use this example in order to check the accuracy of the pole positions and residues obtained in Sect. III.

The four ICSs are given by the partial wave sums
\begin{eqnarray}\label{21}
\sigma_{nn'}(E)= \frac{\pi}{k_{n'}^2}\sum_{J=0}^{\infty}(2J+1)|\delta_{nn'}-S_{nn'}|^2,\q
\br  n,n'=1,2\q\q\q
\end{eqnarray}
where $k_{n'}$ is the wave vector in the incoming channel, and $\delta_{ij}$ is the Kronecker delta.
The PWS (\ref{21}) can be separated into the resonance and background contributions, 
 \begin{eqnarray}\label{21a}
\sigma_{nn'}(E)=\sigma^{res}_{nn'}(E) +B_{nn'}(E),
\end{eqnarray}
where the resonance term is given by the Mullholland formula \cite{Mac}, \cite{S1}, \cite{SA}

 \begin{eqnarray}\label{22}
\sigma^{res}_{nn'}(E)= \frac{8\pi^{2}}{k_j^2} Im \frac{\bar{\lambda}\rho_{nn'}[S^*_{nn'}(\bar{\lambda}^*)-\delta_{nn'}]}{1+\exp(-2\pi i \bar{\lambda})},
\end{eqnarray}
$\lambda=J+1/2$ and a $^*$ denotes complex conjugation.

Exact composition of the background term $B_{nn'}(E)$ is described elsewhere \cite{Mac,S1}, 
its essential property being  smooth behaviour in the region where an ICS is affected by the resonance. Figure 3 shows the residues trajectories (curves $Im \rho_{ij}$ vs. $Re \rho_{ij}$
first introduced in Ref. \cite{RESTRAJ}) evaluated for the Regge trajectory (I) in Fig. 2. 
  \begin{figure}[ht]
\centerline{\includegraphics[width=8cm, angle=-0]{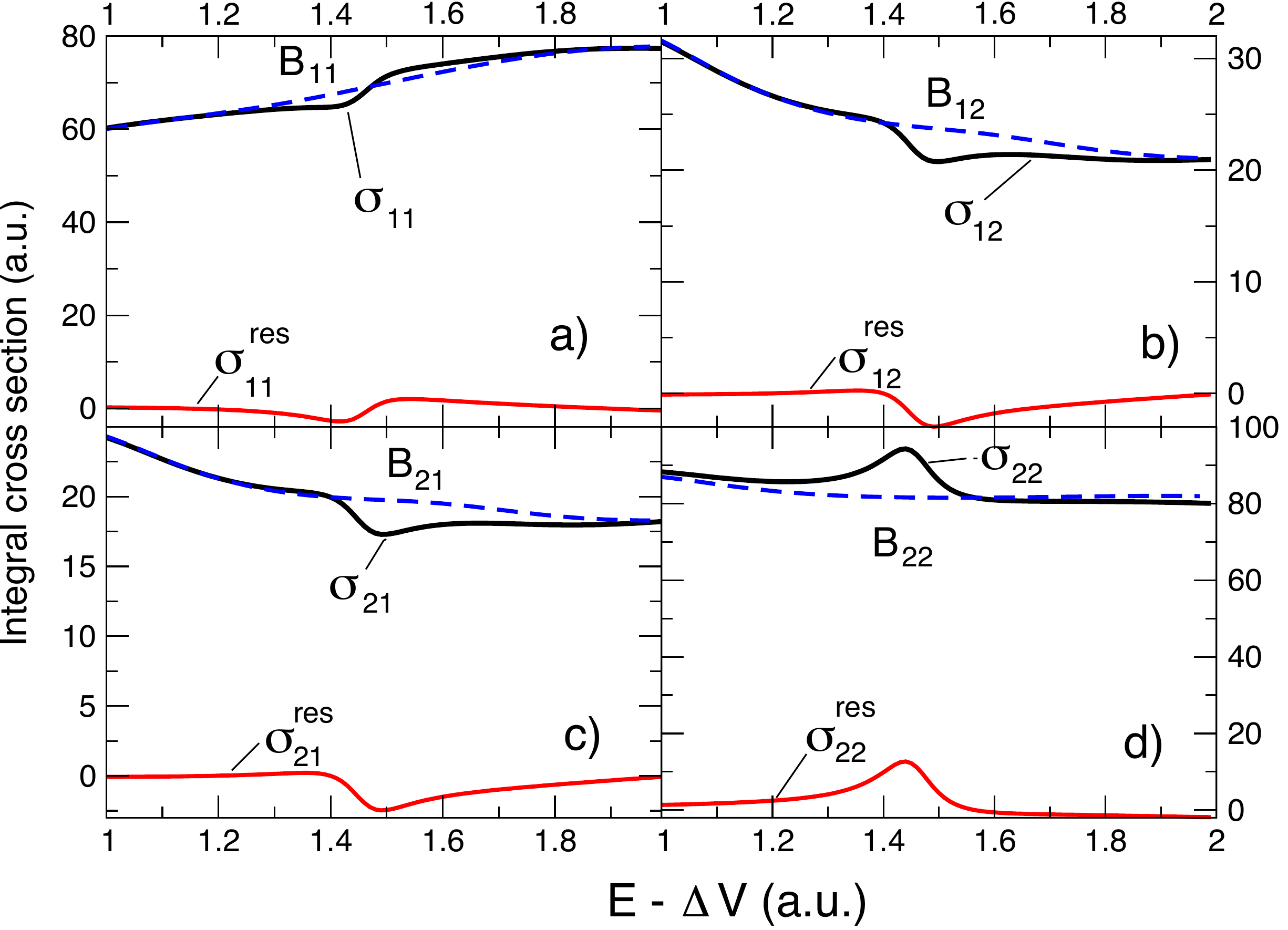}}
\caption{ (Color online) State-to-state integral cross sections $\sigma_{ij}$ vs. $E$ (thick solid)
for (a) i=1, j=1; (b) i=1, j=2; (c) i=2, j=1; (d) i=2, j=2. Also shown are the resonance contributions
$\sigma_{ij}^{res}$ (solid) and the background terms $B_{ij}$ for the trajectory (I) in Fig. 2.}.
\label{fig:1} 
\end{figure}
The full ICSs given by the PWS (\ref{21}), the resonance term (\ref{22}), and the background term obtained as the difference between the two, are shown in Fig.4. The resonance terms $\sigma^{res}_{nn'}(E)$ account for most of the resonance structure at $E\approx 1.46$ a.u. We note that the trajectory (I) in Fig. 2 originates, at low angular momenta, from a bound rather than a metastable state of the two coupled wells. Thus, as in the case of proton impact on neutral atoms \cite{Mac} and also electron-atoms collisions \cite{AZM}, one can expect Eqs.(\ref{21a})-(\ref{22}) to provide an efficient separation of the resonance contribution and to probe important physics.
Discussion of the distinction between two types of trajectories and a modification of the Mulholland formula can be found in Ref.\cite{SA}. Extension of the approach of Ref. \cite{SA} to a multichannel case will be given elsewhere.

\section{Conclusions and discussion}
In conclusion, we advocate a direct method for calculating  Regge pole positions and residues, suitable for systems with a relatively small number of channels.  The method is applied to a simple model designed to mimic electron-atom scattering at energies between the first and the second excitation thresholds. It is shown that inter-channel coupling splits degenerate  Regge trajectories into ones approximately corresponding to the two adiabatic potentials.
Beyond the adiabatic approximation, non-adiabatic effects are seen to be responsible for self intersection of the trajectory $(II)$ shown in Fig.2. The effect of loop formation  has not, to our knowledge, been observed in single channel scattering. 

Finally, the simple model developed here can be improved e.g., by a careful choice of the potential matrix or by including, if necessary, additional $J$-dependent terms in Eq.(\ref{1}). The possibility to use the method for a more accurate description of inelastic electron-atom scattering will be discussed in our future work.
Suffice it to say that this development promises a powerful approach to low-energy scattering with the possibility to probe Regge resonances and the Feshbach resonances occurring in 
Bose-Einstein condesates .

\begin{acknowledgements}
One of us (DS) acknowledges support by the Basque Goverment grant IT472
and MICINN (Ministerio de Ciencia e Innovacion) grant FIS2009-12773-C02-01.
AZM and ZF are supported by US DOE and US AFOSR grants.
DS is also grateful to Karl-Eric Thylwe for useful discussions, suggestions 
and hospitality during visit to Uppsala in May 2010.
\end{acknowledgements}

\end{document}